\documentclass[12pt]{iopart}
\usepackage{graphicx}
\begin{document}

\bibliographystyle{unsrt}

\title{Investigation of a single-photon source based on quantum interference}
\author{T.B. Pittman* and J.D. Franson*}
\address{University of Maryland, Baltimore County, Baltimore, MD 21250}
\author{B.C. Jacobs}
\address{Johns Hopkins University,
Applied Physics Laboratory, Laurel, MD 20723}

\date{\today}

\begin{abstract}
We report on an experimental investigation of a single-photon source based on a quantum interference effect first demonstrated by Koashi, Matsuoka, and Hirano [Phys. Rev. A {\bf 53}, 3621 (1996)]. For certain types of measurement-based quantum information processing applications this technique may be useful as a high rate, but random, source of single photons.
\end{abstract}

\pacs{03.67.Lx, 42.65.Lm, 42.50.Dv}

\maketitle

\vspace*{.25in}
\section{Introduction}
\label{sec:introduction}

Measurement-based quantum logic gates form the basis of Linear Optics Quantum Computing (LOQC) \cite{knill01}, as well as many smaller-scale applications of quantum information processing with single photons. At the present time, one of the key requirements for the realization of these gates is the development of reliable sources of single photons in pure states. In this brief paper, we investigate the potential of a single-photon source based on quantum interference effects.

Specifically, we consider the antibunched light beam that can be generated by mixing a weak coherent state with a phase-locked parametric down-conversion (PDC) source on a 50/50 beam splitter, as was first demonstrated by Koashi, Matsuoka, and Hirano (KMH) \cite{koashi96}.  We describe experimental work that follows up on the original KMH effort by making three main technical improvements: a non-collinear geometry for noise reduction, femtosecond pulses and narrowband filters for improved temporal indistinguishability, and single-mode fibers for spatial mode-matching. This results in a high quantum interference visibility of roughly 80\%, which corresponds to a suppression of the two-photon term by a factor of 2.7 compared to a weak coherent state.

Earlier theoretical analyses by Ou {\em et.al.} have already shown how such a state could be useful in extending the range of quantum communication systems \cite{lu05} and the generation of three-photon entangled states \cite{shafiei04}. Here we investigate the possibility of using this type of antibunched beam as a (random) single-photon source in certain measurement-based quantum information applications. We discuss a specific example that shows how two of these sources could be used to provide frequency un-entangled photons to serve as the input qubits in coincidence-basis two-photon controlled-NOT (CNOT) gates.

\section{Overview}
\label{sec:KMH}

\begin{figure}[t]
\begin{center}
\includegraphics[angle=-90,width=4in]{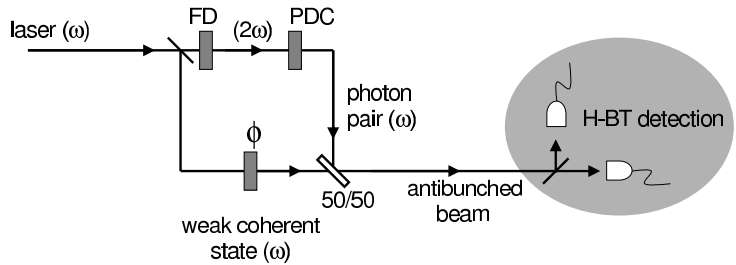}
\end{center}
\vspace*{-1.5in}
\caption{Simplified overview of the KMH technique for generating antibunched light \protect\cite{koashi96}. FD is a frequency doubler and PDC is a parametric down-conversion crystal. The weak coherent state and phase-locked PDC source are mixed at a 50/50 beam splitter. The resulting antibunched beam is observed in a standard Hanbury-Brown Twiss (HB-T) arrangement shown in the shaded region.}
\label{fig:overview}
\end{figure}

A simplified overview of the KMH quantum interference technique \cite{koashi96} for generating antibunched light is shown in Figure \ref{fig:overview}. A small fraction of a pulsed laser beam at frequency $\omega$ is used as the weak coherent state. The remainder of the original laser is frequency doubled to $2\omega$, and then used to pump a PDC crystal which generates pairs of photons back at the fundamental frequency $\omega$. The weak coherent state and PDC source are then mixed at a 50/50 beam splitter, and the pulse train of interest emerges in one of the output ports.

The origin of the antibunching can be understood by considering the probability of simultaneously finding two photons in this beam. This could arise from an amplitude corresponding to the PDC pair, or from an amplitude corresponding to the two-photon term of the weak coherent state.  If these two amplitudes have equal magnitudes, but are out of phase, the probability of simultaneously finding two photons in the beam of interest will be suppressed. \footnote{ Under these conditions, the absence of pairs in the beam of interest implies that photon pairs from either source have separated at the beam splitter. This coherent process can essentially be thought of as the time-reverse of the well-known Hong-Ou-Mandel effect \cite{hong87}: $|2,0\rangle +|0,2\rangle \rightarrow |1,1\rangle$ (see also \protect\cite{rarity90}).}

Closely related effects have also been observed by using the parametric process itself, rather than a beam splitter, to remove the probability of simultaneously finding two photons in a single beam \cite{koashi93,lu02}. In either case, the end result is a beam which Ou {\em et.al.} have aptly named a ``modified coherent state'' because it resembles a coherent state without the two-photon term \cite{lu05}. Additional work in this area includes both theoretical studies  \cite{stoler74,ritze79,grangier88,dodson93,torgerson97,deng00}, and experiments involving systems similar to Figure \ref{fig:overview} (see, for example,  \cite{smithey93,koashi94,kuzmich00,ourjoumtsev06,pittman06}).

\section{Experiment}
\label{sec:experiment}

\begin{figure*}[t]
\begin{center}
\includegraphics[angle=-90,width=5in]{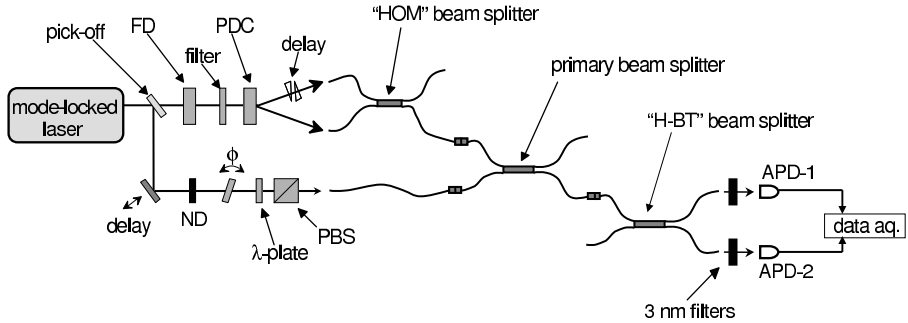}
\end{center}
\vspace*{-2in}
\caption{Schematic of the experimental apparatus using a non-collinear geometry, ultrafast PDC, and single-mode fiber components. Additional details and symbols are described in the text.}
\label{fig:experiment}
\end{figure*}

A schematic of our experimental implementation is shown in Figure \ref{fig:experiment}. The apparatus is a modified version of the one we recently used to generate entangled photon holes \cite{pittman06}. The primary difference between this arrangement and the original KMH experiment is that we use a non-collinear geometry that uses both input ports of the primary 50/50 beam splitter (in analogy with Figure \ref{fig:overview}), whereas the original KMH experiment used a ``single beam''  geometry in which the weak coherent state passed through the PDC crystal, and the output of both sources impinged upon the same port of the primary 50/50 beam splitter.

In addition, we use a non-collinear type-I PDC process and a Hong-Ou-Mandel (HOM) interferometer \cite{hong87} to generate the pairs of PDC photons impinging on the upper port of the primary 50/50 beam splitter.  In contrast to collinear PDC, this arrangement helps eliminate noise photons associated with unblocked UV light from the PDC pumping beam, and residual un-doubled fundamental light from the master laser.

The resulting ``three beam'' arrangement offered a relatively easy method to study the strength of the antibunching effect as a function of the delay between the individual PDC photons (at the HOM beam splitter), as well as the coarse-scale delay and relative phase $\phi$ between the  PDC pair and weak coherent state sources at the primary beam splitter.

The master laser in Figure \ref{fig:experiment}  was a mode-locked Ti:Sapph operating at a repetition rate of 76 MHz, with a fundamental wavelength of 780 nm and a pulse duration of roughly 150 fs. The pulse train was frequency doubled (to 390 nm) in a 0.7 mm thick BBO crystal (labelled FD). The UV pulses pumped a second 0.7 mm thick BBO crystal used for PDC. The 780 nm PDC pairs, as well as the weak coherent state derived from the master laser, were coupled into single-mode 3 dB fused fiber couplers that served as the 50/50 beam splitters. The use of single-mode fibers ensured spatial mode-matching of the independent beams.

The antibunching was observed by using two single-photon detectors (APD-1 and APD-2) in a standard Hanbury-Brown and Twiss (H-BT) type configuration \cite{hanbury56}.  Interference filters with a bandpass of 3 nm (centered at 780 nm) were used before the detectors to ``stretch'' the coherence length of the detected photons. This helped reduce temporal mode-matching problems that originate from pulse-width effects and group velocity dispersion in the PDC process, and are detrimental to the multi-photon interference effects of interest \cite{zukowski95}. The use of a relatively thin PDC crystal with 150 fs laser pulses and 3 nm interference filters represents a different regime than the original KHM experiment, which used a 5.8 mm thick KNbO$_{3}$ PDC crystal, $\sim$100 ps pulses, and an etalon with a bandpass of several GHz.

The magnitude of the weak coherent state was coarsely adjusted by a neutral density filter (ND), and fine-tuned (so that its two-photon component matched the PDC pair production rate) by a rotating half-wave plate and polarizing beam splitter ($\lambda$-plate, PBS). The relative phase $\phi$ between the weak coherent state and PDC source could be adjusted with a tilting glass plate.

\begin{figure}[t]
\begin{center}
\includegraphics[angle=-90,width=3in]{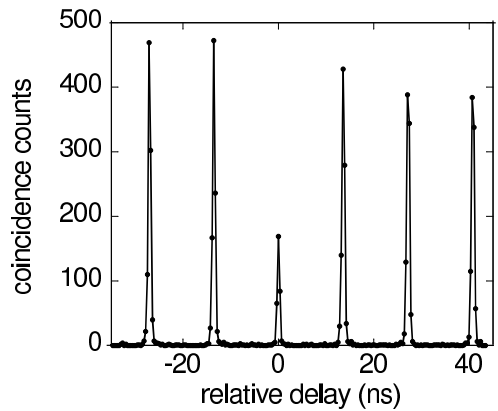}
\end{center}
\vspace*{-.25in}
\caption{Experimental data showing antibunching through quantum interference. The data shows the number of joint detections (between APD-1 and APD-2) per 60 seconds as a function of the difference in detection times. Destructive interference resulted in the central peak being reduced by a factor of 2.7 compared to the average of the side peaks.}
\label{fig:data}
\end{figure}

Data demonstrating the antibunching effect is shown in Figure \ref{fig:data}. The data shows the number of coincidence counts as a function of the relative delay between the two detectors, with $\phi$ set for maximum destructive interference. The integrated area of the peak at zero time delay is 346 counts, while the average of the side peaks is (948.8 $\pm$ 17.7) counts, indicating a suppression of the two-photon term by a factor of roughly 2.7 compared to a weak coherent state.

Analogous data was also obtained with $\phi$ set for maximum constructive interference (eg. bunching). In this run, the value of the central peak was 3,351 counts compared to an average side peak value of (1,002.4 $\pm$ 40.0) counts.  A comparison of the central peaks in antibunched vs. bunched data sets indicates a quantum interference visibility of roughly 80\%.

\section{Single-photon Source for Linear Optics Logic Gates}
\label{sec:results}

In the idealized case of perfect visibility, the two-photon term would be completely suppressed. For certain applications in which higher order terms in the weak coherent state are negligible and/or irrelevant, the KMH technique can therefore be viewed as approximating a source of single-photons: a high-rate stream of optical pulses containing zero or one, but never two, photons \cite{lu05}. Despite its inherently random nature, a single-photon source of this kind may still be useful in certain measurement-based quantum information applications which require detection of the photons of interest.

\begin{figure}[t]
\begin{center}
\includegraphics[angle=-90,width=3in]{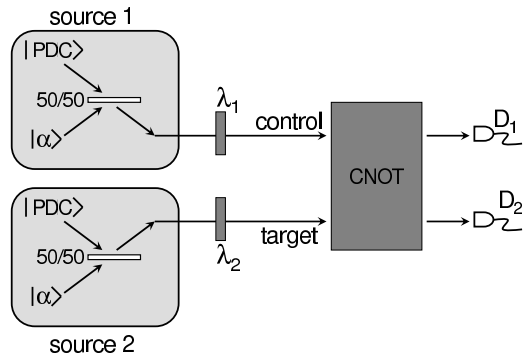}
\end{center}
\vspace*{-.25in}
\caption{Illustration of the use of two quantum-interference-based sources as the inputs to a two-photon coincidence-basis CNOT gate \protect\cite{ralph02,hofmann02,obrien03,3prls05}. In each source, $|PDC\rangle$ and $|\alpha\rangle$ represent, respectively, the PDC and weak coherent state sources of Figure \ref{fig:overview}. Waveplates $\lambda_{1}$ and $\lambda_{2}$ represent polarization-based qubit value preparation, and $D_{1}$ and $D_{2}$ are detectors used for coincidence-basis operation.  In principle, this setup could allow a demonstration of the gate with independent photon sources, but at a data rate comparable to using the signal and idler photons of a single PDC source as the inputs. Although two independent sources are shown, time-sequencing a single source would also be possible.}
\label{fig:cnot}
\end{figure}

Figure \ref{fig:cnot} illustrates an example where two idealized sources of this kind are used as the inputs to an LOQC-type two-photon coincidence-basis CNOT gate of the kind developed in References \cite{ralph02,hofmann02,obrien03,3prls05}. The advantage of using these sources would be the possibility of demonstrating the operation of the gate with independent (frequency un-entangled) photon sources, but at a data rate that would still be comparable to simply using the signal and idler photons from a single PDC source as the inputs.

To get a feel for this argument, we assume we have two similar sources, and use the following number-state-basis notation to describe the weak coherent state and PDC state of each source:

\begin{eqnarray}
|\alpha\rangle \sim |0\rangle +\alpha |1\rangle +\frac{\alpha^{2}}{\sqrt{2}}|2\rangle + \ldots \\
\nonumber
|PDC\rangle \sim |0\rangle  +\frac{\gamma^{2}}{\sqrt{2}}|2\rangle + \ldots
\label{eq:states}
\end{eqnarray}

\noindent where $|\alpha|$ and $|\gamma| \ll 1$, and higher-order terms are ignored. Here the probability (per pulse) of producing a PDC pair is on the order of $|\gamma|^{4}$, and the probability (per pulse) of two photons in the weak coherent state is on the order of $|\alpha|^{4}$. The antibunching effect of interest requires the magnitude of the weak coherent state to be adjusted so that its two-photon amplitude matches that of the PDC source (eg. $|\alpha|^{4} \sim |\gamma|^{4}$).

In this case, each source would emit single photons at a rate proportional to $|\alpha|^{2}$, and the overall data rate of the CNOT gate experiment would therefore be proportional to $|\alpha|^{2} \times |\alpha|^{2} \sim |\gamma|^{4}$. In contrast, note that using single-photon inputs heralded from two independent PDC sources would provide a much smaller data rate proportional to $|\gamma|^{8}$.

Such a demonstration with frequency un-entangled inputs would closely simulate what would be required in using these gates, for example, to build up larger cluster-states from independent entangled pairs. We note, however, that the favorable data-rate-scaling does not, in general, apply to LOQC logic gates involving larger numbers of photons. For example, attempting to use three independent sources in a 3-photon coincidence-basis CNOT gate (of the kind in reference \cite{pittman03}) would lead to undesired events originating from the three-photon component of the weak coherent states.

\section{Summary}
\label{sec:summary}

Quantum interference techniques can offer a relatively robust method for producing antibunched light. In this paper, we have described an implementation of the KMH quantum interference technique by using a non-collinear geometry and single-mode fibers, and working in the regime of femtosecond pulses and narrowband spectral filtering \cite{zukowski95}. Following the arguments of reference \cite{lu05}, the experimentally observed reduction of the two-photon term by a factor of 2.7 (compared to a coherent state) represents a significant step towards a useful source for quantum communications.

In principle, these techniques could also be used to approximate a single-photon source for a limited class of measurement-based quantum information applications. For the situation considered in Figure \ref{fig:cnot}, two sources could allow a high-data rate demonstration of a coincidence-basis CNOT gate with independent photons. However, the experimental challenges in improving the quality of the source demonstrated here to the required level would be significant.

\ack
This work was supported by ARO, DTO, and IR\&D funding.

\noindent *{\em The experimental portion of this work was completed while TBP and JDF were at the Johns Hopkins University Applied Physics Laboratory.}



\end{document}